\documentclass[twocolumn,showpacs,preprintnumbers,amsmath,amssymb]{revtex4}
\usepackage{graphicx}% Include figure files
\usepackage{dcolumn}% Align table columns on decimal point
\usepackage{bm}% bold math
\usepackage{amssymb}

\newcommand\nuh{\nu_h \rightarrow \gamma  \nu }

\newcommand\pair{e^+ e^-}
\newcommand\mix{|U_{\mu h}|^2}

\def\address{\@ifstar{\address@star}%
  {\@ifnextchar[{\address@optarg}{\address@noptarg}}}

\begin{document}

\author{S.N.~Gninenko}
\affiliation{Institute for Nuclear Research of the Russian Academy of Sciences, Moscow 117312}

%\preprint{APS/123-QED}

%\title{Manuscript Title:\\with Forced Linebreak}% Force line breaks with \\
\title{The  MiniBooNE anomaly and heavy neutrino decay}
%\title{The  MiniBooNE anomaly and heavy neutrino decay}

\date{\today}% It is always \today, today,
             %  but any date may be explicitly specified
%\date{June 17, 2009}% It is always \today, today,
             %  but any date may be explicitly specified

\begin{abstract}
The anomaly in the low energy distribution of quasi-elastic neutrino 
events reported by the MiniBooNE collaboration is discussed.
 We show that the observed excess of electron-like events could originate from 
the production and decay of a heavy neutrino ($\nu_h$) in the MiniBooNE 
detector. The $\nu_h$  is created by mixing in  
$\nu_\mu$ neutral-current interactions and  decays radiatively  into 
$ \nu  \gamma$ due to  a transition magnetic moment
between the $\nu_h$ and a light neutrino $\nu$. The energy  measured 
in the detector arises from the subsequent conversion of the decay
photon into  a $\pair$ pair  within the detector volume. 
 The analysis of the energy and angular 
distributions of the excess events suggests that the $\nu_h$ has a mass 
around 500 MeV and the lifetime 
$\tau_{\nu_h} \lesssim 10^{-9}$ s. Existing 
 experimental data are found to be consistent with a
mixing strength between the $\nu_h$ and the $\nu_\mu$ of 
$|U_{\mu h}|^2 \simeq (1-4)\times 10^{-3}$ and  a   
$\nu_h$ transition magnetic moment of $ \mu_{tr} \simeq (1- 6)\times 10^{-9} \mu_B$.     
Finally,  we discuss the reason why no significant excess of low energy 
events has been observed in the  recent antineutrino data.

\end{abstract}

% PACS, the Physics and Astronomy
                             % Classification Scheme.
%\keywords{Suggested keywords}%Use showkeys class option if keyword
                              %display desired

\maketitle

%%%%%%%%%%%%%%%%%%%%%%%%%%%%%%%%%%%%%%%%%%%%%%%%%%%%%%%%%%%%%%%%%%%%%%%%%%%%%%%

%\section{Introduction}

The MiniBooNE collaboration, which was searching for  the  LSND  signal
from neutrino oscillations at FNAL, 
has observed an excess of low energy electron-like events
in charge-current quasi-elastic ($CCQE$) neutrino events 
 over the expected standard  
 neutrino interactions  \cite{mb1}.  This anomaly 
has been  recently confirmed by the finding of more excess events \cite{mb2}. 
While the collaboration has not yet clarified 
the origin of the excess, several models involving new physics   were 
considered to explain the discrepancy \cite{mod}. 

In this work we show that the excess could be explained by 
 the production and decay of a  heavy neutrino ($\nu_h$).
Such  type of neutrinos are present in many interesting extensions of 
the Standard Model, such as 
GUT, Superstring inspired models, Left-Right Symmetric models and others. 
The massive neutrino decays were also considered to explain the LSND signal
\cite{lsnd}. 

The neutrino weak flavor eigenstates
($\nu_e,~\nu_{\mu},~\nu_{\tau},...$) can be different from the mass
eigenstates ($\nu_1,~\nu_2,~\nu_3,~\nu_4...$), but they are related to them, in general,
through a unitary transformation.   A generalized mixing:
\begin{equation}
\nu_l= \sum_i U_{li} \nu_i;~~~l=e,\mu,\tau,...,~i=1,2,3,4,...
\label{mixing}
\end{equation}
results in neutrino oscillations when the mass differences are small,
and in neutrino
decays when the mass differences are large.
If the $\nu_h$  exists, it could be a component of $\nu_\mu$, and, 
as follows from Eq.(\ref{mixing}), it would  be produced by any source of 
$\nu_\mu$ according to the proper mixing $|U_{\mu h}|^2$ and 
 kinematic constraints. The $\nu_h$  can be Dirac or Majorana type \cite{dirac}
 and  
can decay radiatively into $\nu \gamma$, if there is    
 a non-zero transition magnetic moment between the $\nu_h$ and 
a light neutrino $\nu_i$ \cite{moh}.

The MiniBooNE detector is described in details in Ref. \cite{mbdet}. 
It is using an almost pure  $\nu_\mu$ beam originated from
the  $\pi^{+}$ decays in flight, which are  generated 
  by 8 GeV protons from the FNAL booster.
The detector consists of a target, which is  a
12.2 m diameter sphere filled with 800 t of mineral oil, 
surrounded by an outer  veto region. 
The Cherenkov light rings generated by  muon, electron and converted photon 
tracks are used for the reconstruction of the events. The resolutions 
reached on the vertex position, the outgoing particle 
direction and the visible energy are 20 cm, 4 degrees and 
12\%, respectively for $CCQE$ electrons \cite{mbreco}.
The $\nu_\mu$ beam is peaked  
around $\sim 600$ MeV, has a mean energy of  
 $\sim 800$ MeV and a high energy tail up to $\sim$ 3 GeV \cite{mbbeam}.

An excess of $\Delta N=$128.8$\pm 20.4 \pm 38.3$ electron-like events 
 has been observed in the data   
accumulated with $6.64\times 10^{20}$ protons on target.
For the following discussion several distinctive features of the excess events
are of importance \cite{mb2}: a) the excess is  observed for single track events, 
originating either from an 
electron, or from  a photon converted into a $\pair$ pair with a typical 
opening angle $\simeq m_e/E_{\pair}< 1$ degree (for
 $E_{\pair} > 200$ MeV), which is 
too small to be resolved into two separate Cherenkov rings (here,
$m_e, E_{\pair}$ are the electron mass and the $\pair$ pair energy);
b) the reconstructed neutrino 
energy is in the range $200 < E^{QE}_\nu < 475$ MeV, while 
there is no significant 
excess for the region $E^{QE}_\nu > 475$ MeV. 
The variable 
 $E^{QE}_\nu$  is calculated under 
the assumption that the observed  electron track originates from {\bf a} $\nu_e$
$QE$  interaction;  c) the visible 
energy $E_{vis}$   is in the  narrow  region 
$200\lesssim  E_{vis} \lesssim 400$ MeV for events with $E^{QE}_\nu > 200$ MeV;
d) the angular distribution of 
the excess events with respect to the incident neutrino direction 
is wide and consistent with the shape expected 
from $\nu_e CC$ interactions. 
\begin{figure}[h]
\begin{center}
    \resizebox{6cm}{!}{\includegraphics{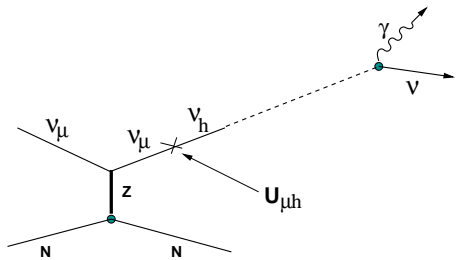}}
     \caption{ Schematic illustration of the $NCQE$ production and  decay of heavy 
neutrino. }
\label{diag}
\end{center}
\end{figure}
To satisfy the criteria a)-d), we propose that  the excess events are  
originated from the decay of a heavy neutrino $\nu_h$. 
The $\nu_h$'s  are produced by mixing in  
$\nu_\mu$ neutral-current ($NC$) $QE$ interactions and  
deposit their  energy via the 
visible decay mode  $\nu_h \to \nu \gamma$,  as  
shown in Fig.\ref{diag}, with the subsequent conversion of the decay
photon into $\pair$ pair in the MiniBooNE target. To make a quantitative estimate, 
we performed simplified simulations of the production 
and decay processes shown in Fig.\ref{diag}.
In these simulations  we used a $\nu_\mu$ energy spectrum parametrized from the  
reconstructed $\nu_\mu CCQE$ events \cite{mbbeam}. 
Since in the MiniBooNE experiment $\nu_h$'s decay over an average distance of 
$\lesssim$ 5 m from the production vertex, the sensitivity is 
restricted to the mass range 100 - 600 MeV 
and to   $\nu_h$ lifetimes  $\tau_{\nu_h} < 10^{-7}$ s. 

\begin{figure}[h]
\begin{center}
    \resizebox{8cm}{!}{\includegraphics{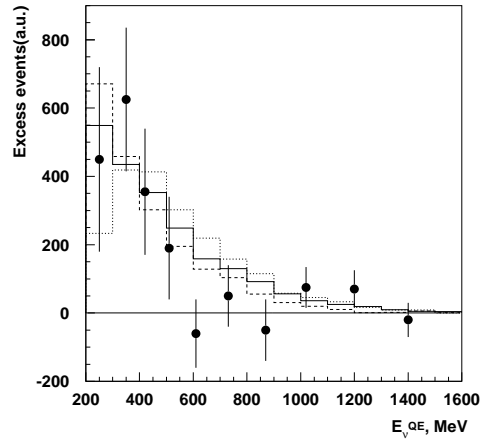}}
     \caption{ Distributions of the excess events from the $\nuh$ decay
 reconstructed as $\nu_e CC$ events as a function of $E^{QE}_\nu$ for 
$|U_{\mu h}|^2=1.5\times 10^{-3}$ and 
a) $m_{\nu_h} = 400$ 
and $\tau_{\nu_h}=3\times 10^{-8}$ s $(\mu_{tr}=2\times 10^{-10}\mu_B)$  (solid); b) $m_{\nu_h} = 400$ 
and $\tau_{\nu_h}= 10^{-10}$ s $(\mu_{tr}= 3\times 10^{-9}\mu_B)$  (dashed); c) $m_{\nu_h} = 600$ 
and $\tau_{\nu_h}=3\times 10^{-8}$ s (dotted).
The dots are experimental points
for the excess events in the MiniBooNE detector. Error bars 
include both statistical and systematic errors \cite{mb2}. 
The  comparison of the distributions with the experimental data
yields a $\chi^2$ of 10.2, 11.2, and 17.2 for 
8 DF corresponding to 27\%, 24\%, and $\simeq$ 5\% C.L. for a),b) and c), respectively. }
\label{sp1}
\end{center}
\end{figure}

\begin{figure}[h]
\begin{center}
    \resizebox{8cm}{!}{\includegraphics{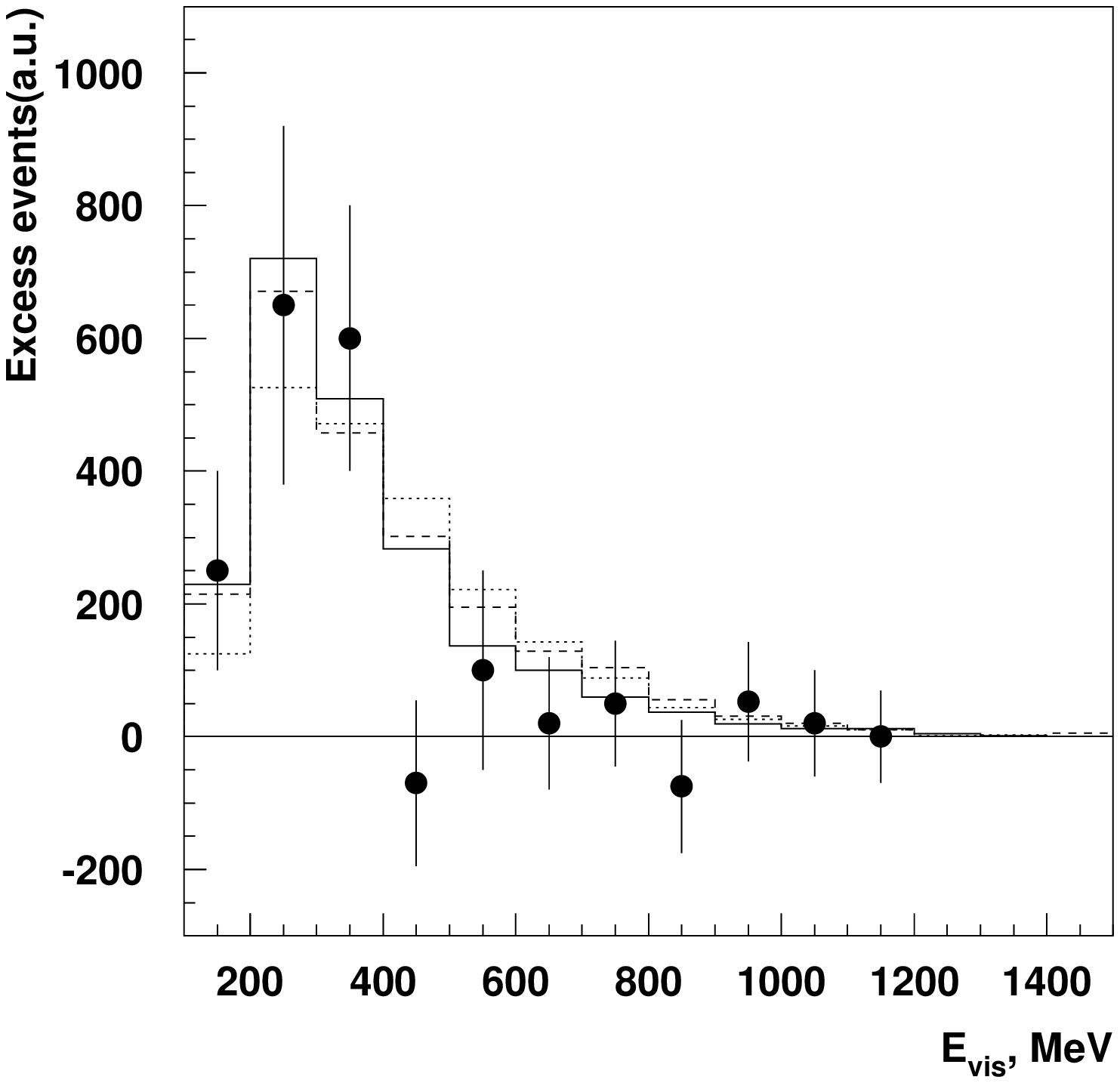}}
     \caption{ Distributions of the excess events from the $\nuh$ decay
 reconstructed as $\nu_e CC$ events as a function of $E_{vis}$ for 
$E^{QE}_\nu > 200$ MeV.
The  comparison of the distributions with the experimental data
yield a $\chi^2$ of 9.7, 10.3, and 16.8 for 
8 DF corresponding to 28\%, 27\%, and $\simeq$ 5\% C.L. for a), b) and c), respectively.
The legend is the same as in Fig.\ref{sp1}.}
\label{sp2}
\end{center}
\end{figure}

\begin{figure}[h]
\begin{center}
    \resizebox{8cm}{!}{\includegraphics{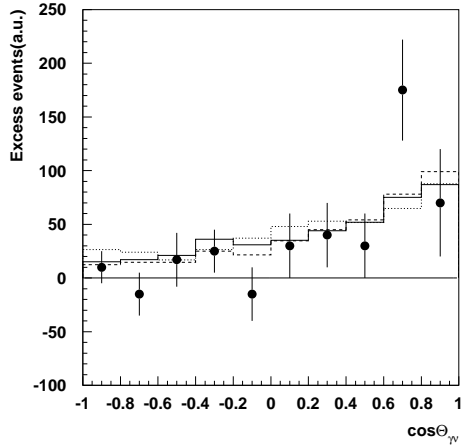}}
     \caption{ Distribution of the excess events from the $\nu_h$ decay
 reconstructed as $\nu_e CC$ events 
as a function of  $\cos \Theta_{\gamma \nu}$ for 
$300< E^{QE}_\nu < 400$ MeV.
The  comparison of the distributions with the experimental data
yields a $\chi^2$ of 11.6, 11.1, and 15.6 for 
8 DF corresponding to 23\%, 24\%, and $\simeq$5\% C.L. for a), b) and c), respectively.
The  notations are the same as in Fig.\ref{sp1}.}
\label{cos}
\end{center}
\end{figure}

In Fig.\ref{sp1}-\ref{cos} the distributions of 
the kinematic variables $E^{QE}_\nu, E_{vis}$ and $\cos (\Theta_{\gamma \nu})$ 
for the $\nuh$ events are shown 
for  $m_{\nu_h}=400$ and 600 MeV and 
$\tau_{\nu_h}=3\times 10^{-8}$ and $10^{-10}$s.  
These distributions were obtained assuming 
 that the  $\pair$ pair from the converted photon 
 is mis-reconstructed  as a  single track from the $\nu_e QE$ reaction.

Simulations are in reasonable agreement with the experimental distributions.  
For instance, for the distribution
 shown in Fig.\ref{sp1}, the
comparison with MiniBooNE  data
yields a $\chi^2$ of 10.2 (17.2) for 8 DF corresponding to 27\% ($\simeq 5\%$) CL. 
for $m_{\nu_h} =400~(600)$ MeV and $\tau_{\nu_h}=3\times 10^{-8}$ s. 
The simulated excess events, shown in Fig.\ref{sp2},
are  mainly distributed  in the
narrow region $ 200 \lesssim E_{vis} \lesssim 400$ MeV.
The fraction of events in the region  $200 <  E_{vis} < 400$ MeV 
is $\sim 70\%$. 
The remaining  events are distributed over the region 
$400\lesssim E_{vis} \lesssim 1200$ MeV, where they can be  hidden by the low 
statistics. 
%to identify an excess
%is not easy due to the low statistics.
The simulations showed that the shape of the $E^{QE}_\nu$ and $ E_{vis}$  
distributions 
is sensitive to the choice of the $\nu_h$ mass and lifetime: the   
shorter the $\nu_h$ lifetime the broader the visible energy 
spectrum.  The best 
fit results suggest  that the $\nu_h$ mass 
 is in the region $200 \lesssim m_{\nu_h}\lesssim 600$ MeV and the 
lifetime is $\tau_{\nu_h} < 10^{-7}$ s.
The estimate of the mixing parameter $|U_{\mu h}|^2$ was 
performed by using the following relations.
For a given flux of heavy neutrinos, $\Phi(\nu_h)$,  the expected number  
 of the decays in the MiniBooNE detector is given by 
$\Delta N =\int \Phi(\nu_h) P_{dec} P_{conv} \epsilon dE_{\nu_h} dV$,
where $ P_{dec}$ and  $P_{conv}$  are the 
  probabilities of the $\nu_h$ decay and the photon conversion in the detector, 
$\epsilon $ is the  overall detection  efficiency, 
 and the integral is taken over the detector fiducial volume. 

The flux  $\Phi(\nu_h)$ was estimated from the expected 
number of the  $\nu_\mu NC$  events 
   times the mixing 
$|U_{\mu h}|^2$, taking into account the threshold effect due to the heavy 
neutrino mass. The total number of 
reconstructed $\nu_\mu CC$ events in the detector \cite{mbbeam} was used for 
normalization.
The probability  of the heavy neutrino to decay radiatively 
in the fiducial volume at a distance $r$ from the primary vertex is given by 
$P_{dec}=[1-\exp (\frac{-rm_{\nu_h}}{p_{\nu_h} \tau_{\nu_h}})]\frac{ \Gamma (\nuh)}{\Gamma_{tot}}$, where the last term is the branching fraction $Br(\nuh) \simeq 1$ (see below).
Taking into account  the 
ratio $\nu_\mu NCQE / \nu_\mu CCQE \sim $ 0.43,  
the number of  $\nu_\mu CCQE$ events observed 
\cite{mb1,mb2} and assuming that almost all $\nuh$ decays occur inside 
the fiducial volume of the detector, we estimate the $\mix$ to be in the range 
\begin{equation}
|U_{\mu h}|^2 \simeq (1-4)\times 10^{-3}.
\label{mixlim}
\end{equation}
This result is mainly defined by the uncertainty on  
the number of excess events. Eq.(\ref{mixlim}) is valid 
for  the mass region 
$400 \lesssim m_{\nu_h} \lesssim  600$ MeV. The lower limit is 
set to 400 MeV to avoid stringent constraints on  
$|U_{\mu h}|^2$ for the mass region $ m_{\nu_h} \lesssim 400$ MeV 
from experiments searching for a peak from $\pi, K \to \mu + \nu_h$ 
decays \cite{pdg}. 
The $\nu_h$ lifetime  due to a transition 
moment $\mu_{tr}$ is given by \cite{moh}
\begin{eqnarray}
\tau_{\nu \gamma}^{-1}=\frac{\alpha}{8}\bigl(\frac{\mu_{tr}}{\mu_B}\bigr)^2 \bigl(\frac{m_{\nu_h}}{m_e}\bigr)^2 m_{\nu_h}
\label{rate}
\end{eqnarray}
and for $\tau_{\gamma \nu} < 10^{-7}$ s results in 
$\mu_{tr} > 10^{-10}{\mu_B}$. The total $\nu_h$ decay width is
$\Gamma_{tot} = \Gamma(\nu_h \to \nu \gamma)+ \Sigma \Gamma_i$,
where $\Gamma(\nu_h \to \nu \gamma)$ is the $\nuh$ decay rate, and $\Sigma \Gamma_i$ is 
the sum over  decay modes whose decay rate is 
  proportional to the square of the mixing $|U_{\mu h}|^2$. 
The dominant contribution to  $\Sigma \Gamma_i$ comes from   
 $\nu_h \to\nu_\mu ee, \nu_\mu \pi^0, \nu_\mu \nu\nu, \mu \pi$ 
decays,  for which the rate calculations can be found, e.g. in \cite{gorby}.
For $m_{\nu_h} \simeq 500$ MeV and $\mu_{tr} > 10^{-10}{\mu_B}$, 
we found that 
the radiative decay is dominant, $Br(\nu_h \to \gamma \nu)>0.5 $.
For example, 
 for $\mu_{tr}= 10^{-9}\mu_B$ and  $|U_{\mu h}|^2=1.5\times 10^{-3}$, 
the expected ratio of decay rates for $\gamma \nu~ :~ \mu \pi~ : ~e \mu \nu~ :~ \mu \mu \nu$ is 0.984 : 0.011 : 0.0016 : 0.00067. 

One may wonder if the mixing strength of 
  Eq.(\ref{mixlim}) is consistent with the results of 
previous searches for $\nu_h$ decays. 
The $\nu_h$ mass region around 500 MeV was covered by many
experiments \cite{gorby,pdg,atre}. 
However, 
%all these experiments searched for 
%$\nu_h$'s decays into charged particles in the final state. 
none of  these experiments has reported  a bound on the mixing 
strength $|U_{\mu h}|^2$ or on the combination $|U_{\mu h}|^2 \mu_{tr}$,
 for the radiative $\nuh$ neutrino decay.
The best limit $|U_{\mu h}|^2 \lesssim 10^{-6}$ for the mass region $m_{\nu_h}\simeq 500$ MeV 
 was derived from a search for  $\nu_h \to \mu \pi, \mu \mu \nu , \mu e \nu  $ decays  in the  NuTeV beam dump experiment \cite{nutev}  (see also  \cite{bebc,charm2, mishra}). 
It was assumed that these decay modes are dominant and 
 that the $\nu_h$ is a relatively long lived particle, 
i.e $\frac{L m_{\nu_h}}{p_{\nu_h} \tau_{\nu_h}} \ll 1 $, where $L \simeq 1.4\times 10^3$ m is the distance 
between the target and the detector.  
Consider now our case with
 $|U_{\mu h}|^2 =(1-4)\times 10^{-3}$, $m_{\nu_h} = 500$ MeV and  
$\mu_{tr} = 10^{-9} \mu_B$. This gives the $\nu_h$ 
lifetime $\tau_{\nu_h}= (1.5-1.4)\times 10^{-9}$ s. Due to the 
larger mixing 
the $\nu_h$ flux at the target would increase
by a factor $\simeq (1-4)\times 10^3$. However, taking into account
the  attenuation of the flux due to the  rapid decay of $\nu_h$'s, the total 
number of signal events in NuTeV  would  decrease
 by a factor $(10-3)$ compared to the number of events expected  
for a long lived $\nu_h$'s produced and decaying through the mixing 
$|U_{\mu h}|^2 = 10^{-6}$.  
In this estimate the average $\nu_h$ momentum is 
$<p_{\nu_h}>\simeq 100$ GeV and the decay region length is $l=34$ m \cite{nutev}. Finally,  we find that for 
\begin{equation}
\mu_{tr} \gtrsim 10^{-9} \mu_B
\label{nutev}
\end{equation}
the NuTeV limit is  not constraining mixing of Eq.(\ref{mixlim}).
Note that a short $\nu_h$ lifetime is also necessary to avoid
the constraints coming from 
cosmological and astrophysical considerations \cite{dolg2}.

The best limit for the large mixing (short lifetimes),  $|U_{\mu h}|^2 \lesssim 10^{-3}$ for the masses
around 500 MeV, was derived 
by CHARM-II from a search for  the $\nu_h$
 production and $\nu_h \to \mu \mu \nu$ decays within their detector \cite{charm2}.
Taking into account the $\nuh$ decay, the $<p_{\nu_h}>\simeq 24$ GeV and 
$l=35$ m, we find that the number of the expected $\mu \mu \nu$ signal events 
 in CHARM-II is  $N_{\mu\mu\nu}\simeq 0.46\times 10^{-12} \frac{|U_{\mu h}|^4}{(\mu_{tr}/\mu_B)^2}$.
For $\mu_{tr}\geq 3\times  10^{-9} \mu_B$ and for mixing of Eq.(\ref{mixlim})
this results in  $N_{\mu\mu\nu} < 0.05-0.8$  events. To evade the CHARM-II
 limit
 for the region  
$10^{-9} < \mu_{tr} < 3\times 10^{-9} \mu_B$  the mixing is required to be 
in a slightly more restricted range 
 $10^{-3}< |U_{\mu h}|^2 < 1.5 \times 10^6 \frac{\mu_{tr}}{\mu_B}$ 
compare to  that of Eq.(\ref{mixlim}). For example,
 for $\mu_{tr}=2\times 10^{-9}\mu_B$ the allowed range is 
 $10^{-3}< |U_{\mu h}|^2 < 3 \times 10^{-3}$. Thus,  we see that 
most of the allowed  $(\mu_{tr};|U_{\mu h}|^2)$ parameter 
space corresponding to Eqs.(\ref{mixlim},\ref{nutev}) is not  constrained
by the CHARM-II limit.  
%However, taking into account the $\nuh$ decay,  this results into  a 
%weaker limit $|U_{\mu h}|^2 \lesssim 0.01$  for our case.

Consider now bounds from LEP experiments \cite{pdg}.
For the mass region around 500 MeV, 
the model independent limit from  the  searches for 
the $Z \to \nu \nu_h$ decay is $|U_{\mu h}|^2 \lesssim 10^{-2}$, 
(see e.g. \cite{delphi}) which is compatible with Eq.(\ref{mixlim}). 
Direct searches for radiative decays of an excited 
neutrino $\nu^* \to \gamma \nu$ produced in $Z\to \nu^* \nu$ decays have 
been also performed \cite{pdg}.  The best limit from ALEPH  is
$Br(Z\to \nu \nu ^*) Br(\nu* \to \gamma \nu) < 2.7\times 10^{-5}$ \cite{aleph}.
As the  experimental signatures for the 
$\nu* \to \gamma \nu$ and $\nuh$ decays
are the same, we will use this bound for comparison.
 The number of expected $\nuh$
events in ALEPH is proportional to $Br(Z\to \nu \nu_h) Br(\nuh) [1-\exp (-\frac{l m_{\nu_h}}{p_{\nu_h} \tau_{\nu_h}})]$, with $l\simeq 1$ m and $p_{\nu_h}\simeq 45$ GeV. 
Taking into account  
$\frac{Br(Z\to \nu \nu_h)}{Br(Z \to \nu \nu )} \simeq  |U_{\mu h}|^2$
and  using Eq.(\ref{rate}), we find
\begin{equation}
|U_{\mu h}|^2\times \Bigl(\frac{\mu_{tr}}{\mu_B}\Bigr)^2 < 3.5 \times 10^{-20}. 
\end{equation}
Using Eq.(\ref{mixlim}) 
results in $\mu_{tr} \lesssim (6-3)\times 10^{-9} \mu_B $, which is 
consistent with Eq.(\ref{nutev}).
%Thus, we see that LEP bounds constrain the product 
%$\mu_{tr}\times |U_{\mu h}|^2 < 10^{-12} \mu_B$ which is 
%consistent with  Eq.(\ref{mixlim}).
%Thus, we see that LEP bounds are 
%consistent with  Eq.(\ref{mixlim}).

The limit on the $\mu_{tr}$ between the $\nu_h$ and the $\nu_\mu$ 
has been obtained in Ref.\cite{gk1}, based on the idea of the 
Primakoff conversion $\nu_\mu Z \to \nu_h Z$   
of the muon neutrino into a heavy neutrino in the external Coulomb field 
of a nucleus $Z$, with the subsequent $\nuh$ decay.  
By  using the results  from the NOMAD experiment \cite{nomad1, nomad2}, a 
model-independent bound $\mu_{tr}^{\mu h} \lesssim 10^{-8} \mu_B$ was 
set for the $\nu_h$ masses around 500 MeV (see Table 1 and Fig.2 in 
Ref.\cite{gk1}), which is also consistent with  Eq.(\ref{nutev}).  

The low statistics anti-neutrino ($\overline{\nu}_\mu$) data collected by the MiniBooNe 
seem to show no low-energy excess  \cite{antinu}. 
 An analysis of these data within the framework discussed above 
suggests that the excess is not seen due to the lower $\overline{\nu}_\mu$
 energy. Indeed, the $\overline{\nu}_\mu$ flux peaks at $\sim 400$ MeV and has 
a mean energy of $\sim 600$ MeV \cite{mbbeam}. If the $\overline{\nu}_h$ 
mass is  around 500 MeV, the $\overline{\nu}_h$
  production is kinematically suppressed for $\overline{\nu}_\mu$
energies below the mean energy. Instead of the 
expected excess of $\sim$ 40 events \cite{antinu},
 a smaller excess of $\sim 23$ events  is expected in the antineutrino data. 

In summary, we see  that the interpretation of the MiniBooNe anomaly  
 based on the production and  visible decay  of a  heavy  neutrino is  
compatible with all the four  constraints a)-d). The shape of the excess 
events in several 
kinematic variables is found to be consistent with the distributions obtained 
within this interpretation. The reason why the excess is not 
observed in the recent antineutrino data \cite{antinu} 
is clarified. A definite conclusion on the presence of $\overline{\nu}_h \to\overline{\nu}_\mu \gamma$ events can be drawn 
when the $\overline{\nu}_\mu$ statistics is substantially increased.
 Our results  for the 
mixing strength $|U_{\mu h}|^2 \simeq (1-4)\times 10^{-3}$ and for the magnetic moment $\mu_{tr}\simeq (1- 6)\times 10^{-9} \mu_B$    
are compatible  with the results from 
previous experiments. Values of $\mu_{tr}$ larger than $10^{-10} \mu_B$ could be obtained e.g. in the framework of the Zee model \cite{moh}. 
Our analysis gives a correct order of magnitude
for the parameters $|U_{\mu h}|^2$ and  $\mu_{tr}$
 and may be improved by more accurate
and detailed simulations of the MiniBooNE detector, which are
beyond the scope of this work. 
%To clarify 
%the origin of the MiniBooNE anomaly a new dedicated experiment
%with a detector able to distinguish electrons and photons would be crucial.  
We note that an analysis of the excess of events due to 
the $\nuh$ decay may also  be possible with existing neutrino data. 
New results could be obtained   with  NOMAD \cite{nomad1},
SciBooNE \cite{sci} and   K2K near detectors \cite{k2k}, see also \cite{gg}. 
 The author thanks S. Brice, A.D. Dolgov, S.H. Hansen, D.S. Gorbunov, 
 N.V. Krasnikov, V.A. Matveev, V.A. Rubakov, M.E. Shaposhnikov, 
and R. Van de Water for useful discussions and/or comments, and 
M. Kirsanov, R. Petti and D. Sillou for help.
  This work was supported by Grant RFBR 08-02-91007-CERN.


\begin{thebibliography}{99}
\bibitem{dirac}
For definitness we will consider further the first case.
\end{thebibliography}
\end{document}